\def\cm2{cm$^{-2}$}

\def\c2{C~{\sc ii}}
\def\c4{C~{\sc iv}}
\def\fe2{Fe~{\sc ii}}
\def\fe3{Fe~{\sc iii}}
\def\mg1{Mg~{\sc i}}
\def\mg2{Mg~{\sc ii}}
\def\si2{Si~{\sc ii}}
\def\si4{Si~{\sc iv}}
\def\al2{Al~{\sc ii}}
\def\al3{Al~{\sc iii}}
\def\o1{O~{\sc i}}
\def\n1{N~{\sc i}}
\def\h1{H~{\sc i}}

\def\approxlt{\mathrel{\spose{\lower 3pt\hbox{$\sim$}}
        \raise 2.0pt\hbox{$<$}}}
\def\approxgt{\mathrel{\spose{\lower 3pt\hbox{$\sim$}}
        \raise 2.0pt\hbox{$>$}}}

\documentclass[article]{gwp80}
\usepackage{graphicx}
\usepackage{amssymb}
\tabletypesize{\normalsize}  

\def\plotone#1{\centering \leavevmode
\includegraphics[width=.95\columnwidth]{#1}}

\def\plotone#1{\centering \leavevmode
\includegraphics[width=.95\columnwidth]{#1}}

\shortauthors{M. Chadid}
\shorttitle{Atmosphere of RR\,Lyrae Stars}

\begin{document}
\large    
\pagenumbering{arabic}
\setcounter{page}{29}

\title{Atmospheres of RR\,Lyrae Stars\\ \\
\large{Hypersonic Shock waves, Helium line detection and Magnetic fields}\\
\large{A few spectra for George Preston on the occasion of his birthday}}
%
%
\author{{\noindent Merieme Chadid{$^{\rm 1,2}$}\\
\\
{\it (1) Observatoire de la C\^ote d'Azur, Universit\'e Nice Sophia-Antipolis, UMR 6525, \\
\ \ \ \ \ \ \  Parc Valrose, 06108 Nice Cedex 02, France\\
(2) Antarctica Research Station, South Pole TAAF, Antarctica} 
}
}

%
%
\email{(1) chadid@unice.fr}


\begin{abstract}
Although the RR Lyrae stars have been studied for more than a century, they  remain highly amazing in improving our knowledge of the Universe and keeping us up awake 
during even the polar night (Chadid et al. 2010). Here I will report some results of observations of RR\,Lyrae stars that I presented during George Preston Conference to 
celebrate his  birthday. I will not report the RR\,Lyrae polar photometric results and the expeditions to Dome Charlie --South Pole. But, I would like to show how greatly
 my spectroscopic 
and spectro-polarimetric works consistently follow the line of Preston's results. When I started working on RR Lyrae atmospheres, and during  intensive/comparative  studies with 
Preston's spectroscopic works (1961, 1962, 1964 \& 1965) on atmospheric phenomena, I unexpectedly found the existence of hypersonic shock waves (Chadid 1996 \& 2008) in the atmosphere of RR\,lyrae stars 
and how strongly they are connected  with the turbulence amplification mechanism (Chadid 1996b) and the Blazhko modulation (1997). In this work, I will point out how 
the hypersonic shocks  are at the origin of  helium line formation and connected with the helium emission and line doubling detected by Preston (2009, 2011).

Here, I will present new neutral and single ionized helium lines  in  hypersonic Blazhko star S\,Arae. The He II appears as a weak emission and its origin is strongly connected with the 
hypersonic shock in S\,Arae. The shock is extremely strong and can reach a Mach number greater than 30. The HeI line appears first as an emission line and, occurs during a time  interval  
of 2\,\% of the pulsation period, followed by the HeI absorption doubling phenomenon which occurs simultaneously with the neutral metallic absorption line phenomenon during a time interval 
of about 8\,\% of the pulsation period.

Despite the fact that several authors claimed to detect a magnetic field, Preston (1967) reported the absence of any strong magnetic 
field after two years of observations. This is consistent with my recent spectro-polarimetric measurements (2004), over four years, concluding that RR\,Lyr is a bona fide nonmagnetic star,  
leading to the falsification of all the models of the Blazhko effect requiring strong photospheric magnetic fields. Here,  I will report on a new series of high precision circular polarization 
spectroscopic observations of the Blazhko star RV\,UMa.  The longitudinal field measurements of RV\,UMa show a  mean longitudinal magnetic field value of $B_{\ell} = -13\pm 35$~G. None of 
measurements of  RV\,UMa shows a significant detection over a whole pulsation period. Finally, all these spectroscopic  results  reveal a serious challenge  for explaining the complex behavior 
in the pulsation of hypersonic atmosphere of RR\,lyrae stars.

\end{abstract}

\section{Introduction}
Helium \& Hydrogen emissions and double metallic absorption lines  were first observed in W\,Virginis stars   by Sanford in the 1940s. Schwarzschild (1952) invoked the propagation of 
shock waves across the atmosphere to explain the double absorption lines during rising light. The first quantitative model of shock waves explaining the Hydrogen emissions was developed 
by Whitey (1956) as reported by Wallerstein (1959) who rediscovered the emission lines of the He\,I. He\,I emission lines were used to confirm the shock model since they 
indicate that the high temperature behind the shock was sufficient to ionize Helium once, and then Hydrogen. 

In RR\,lyrae stars, the first detection of H--emissions and the double H--absorption lines were reported by Sanford (1949). Iroshnikov (1962) interpreted the observations of metallic and 
hydrogen line radial velocities, splitting of the spectral lines and hydrogen emission lines of RR\,Lyr made by Sanford (1949) with the help of a model in which the shock wave propagates 
through an exothermic gas in the atmosphere of RR\,Lyr. He concluded that the discontinuity in the radial velocity curve, the emission and double absorption hydrogen lines, at the maximum of 
the light curve, would be the consequence of shock waves propagating through the atmosphere of RR\,Lyr.

The best spectroscopic observations had been made half a century ago  by George Preston and his collaborators (Preston 1961 \& 1962; Preston and Paczynski 1964; and Preston et al. 
1965). Using a spectrogram (10.7 \AA mm$^{-1}$) attached to the 3 m telescope of Lick Observatory, they rediscovered  the H--emissions which occurs only during an interval 
of 4\,\% of the period, followed by the  double H--absorption lines during rising light, especially  in RR\,Lyr and X\,Ari. They explained the phenomena as a consequence of the propagation 
of a strong shock wave through the atmosphere of RR\,Lyrae stars. Thanks to the improvements of CDD detectors  compared to photographic plates, Chadid and Gillet (1996a) detected for 
the first time the doubling phenomenon over the ionized metallic absorption lines of RR\,Lyr by using the  ELODIE  spectrograph attached to the 2-m telescope at Observatoire de Haute Provence, France.
  They interpreted 
the phenomenon as the consequence of a ``two--step'' Schwarzschild mechanism. The shock is at first receding from the observer, becomes stationary and then advances. Later, they confirmed the 
metallic line doubling phenomenon in the one of largest amplitude RR\,ab stars, S\,Arae, and they detected for the first time a new mechanism, the neutral absorption line disappearance phenomenon during 
rising light, occurring during an interval of 8\,\% of pulsation (Chadid et al. 2008)  using UVES at VLT-UT2. They interpreted the phenomenon as a consequence of an hypersonic shock existence in the 
photosphere of RR\,Lyrae stars. 

``{\it ... but no one applied Wallerstein's ideas about W\,Virginis to RR\,lyrae stars during the next half century (Observers \& theorists alike). Until, quite by accident... I detect 
emission or/and absorption lines of Helium in  RR\,lyrae spectra...}'' George Preston said in a private communication. Thus, Preston (2009) detected for the first time the He\,I spectrum during 
rising light of RR\,Lyrae stars on a time scale of $\sim$ 2\,h. This discovery adds new constraints on the rapidly evolving temperature structure of the atmosphere towards a better understanding of
hypersonic shock waves in atmosphere of RR\,Lyrae stars (Chadid et al. 2008).

More than a century after its discovery (Blazhko 1907), the Blazhko effect still remains a big challenge. It greatly complicates our accurate measurements of the atmospheric dynamics of RR\,Lyrae stars 
and introduces unavoidable complications in an understanding of stellar pulsation and evolution. George Preston had great adventures with the Blazhko effect, when he tried to understand 
the motions of his favorite star TY\,Gruis (Preston et al 2006a and 2006b). ``{\it ...Blazhko just makes it worse. I dislike Blazhko... So many Blazhko vectors aimed at me make me 
nervous...}'' George said. One of the models, trying to explain the Blazhko effect, predicts the dependence of the
Blazhko amplitude upon the strength of a magnetic field of the order of 1.5\,kG  (the oblique--dipole rotator model). However, from the observational point of view, the question 
whether RR\,Lyrae Blazhko stars are magnetic is still a matter of debate. Until recently, RR\,Lyr itself was the only object that has been the target of spectro-polarimetric observations and 
 with contradictory results.  Babcock (1958) and Romanov et al. (1987, 1994) reported a variable magnetic field with amplitude around 1.5\, kG. Preston (1967) and Chadid et al. (2004) reported 
no
 detection. having covered several pulsation periods and Blazhko cycles spread over a period of 4--years with high precision ($\sigma$\,$\sim$\,80\,G)
longitudinal magnetic field measurements. Chadid et al concluded that RR\,Lyr is a bona fide nonmagnetic star. 

As said by George Preston  ``{\it ... Unlike the redwood trees, when you've seen one RR\,ab, you have not seen them all...}''; thus, in order to evaluate the Preston and Chadid et al. 
non--detection 
of the magnetic field to others RR\,ab Blazhko stars, I have obtained a new series of high precision circular polarization spectroscopic observations of the Blazhko star RV\,UMa that I 
report in this work with the helium line detections in the hypersonic atmosphere of S\,Arae.  

In this paper, the hump and the bump will be used according the adopted nomenclature used by Christy (1966). The hump occurs on the rising branch just 
before the luminosity maximum while the bump appears during decreasing light just before the luminosity minimum. 

\section{Helium line detection, Shocks and hypersonic motions}
\subsection{High resolution spectroscopic observations}\label{obs}
High resolution spectra, presented in this work, are based on the Blazhko RR\,ab star, S\,Arae, one of the largest amplitude RR Lyrae stars, 
with pulsation period of 0.452\,days, Blazhko cycle is 48 days and V--magnitude is between 10 and 11.5. I used the high resolution 
UV--Visual Echelle Spectrograph UVES at the VLT--UT2 (Kueyen) on Cerro Paranal covering the 376--500  and 672--1000 ranges (ESO programs 
programs P67.D-0321(B) and P67.D-0321(A)). A detailed description of observations and data reduction are given in Chadid et al. (2008).
In order to achieve an extended coverage in the wavelength range, I also used the  High-Accuracy Radial Velocity Planetary Searcher Spectrograph 
HARPS at ESO 3.6-m telescope on La Silla Observatory covering the wavelength range from 380 to 690\,nm (ESO program P79.D--0462) with 
more observation details in Chadid et al.(2010). 

\subsection{Shock wave effects }
\subsubsection{Schwarzschild mechanism}
\begin{figure}
\centering
\includegraphics[width=0.5\linewidth]{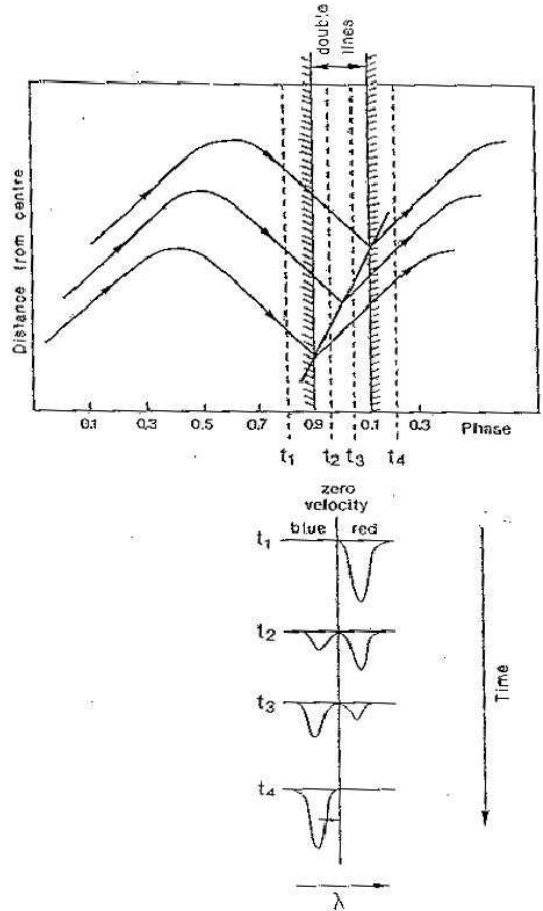}
\vskip0pt
\caption{Schwarzschild's mechanism pointing towards the shock wave propagating in the atmosphere and the absorption line doubling phenomenon.}
\label{schw}
\end{figure}

Spectra of RR\,Lyrae stars show Hydrogen (double) emission lines, Helium emission lines,  neutral metallic line disappearance phenomenon, 
line broadening and doubling phenomena, which cause a discontinuity in the radial velocity curve. All these observational features are 
interpreted by the presence of shock waves propagating  in their atmosphere. Schwarzschild (1953) was the first who
pointed towards the shock stratification as the origin of the double absorption lines. The Schwarzschild's 
mechanism is the temporal sequence followed by the blue shifted and red shifted components of a double absorption line (Figure\,\ref{schw}). 
An upward propagating shock  wave cuts the layer of formation of an absorption line in two parts with opposite motions. The red shifted component
appears first alone and followed by the doubling phenomenon. When the shock is far away from the layer, only the blue shifted 
component is visible. 

\vfill\eject

\subsubsection{Shock wave structure}
\begin{figure}
\centering
\plotone{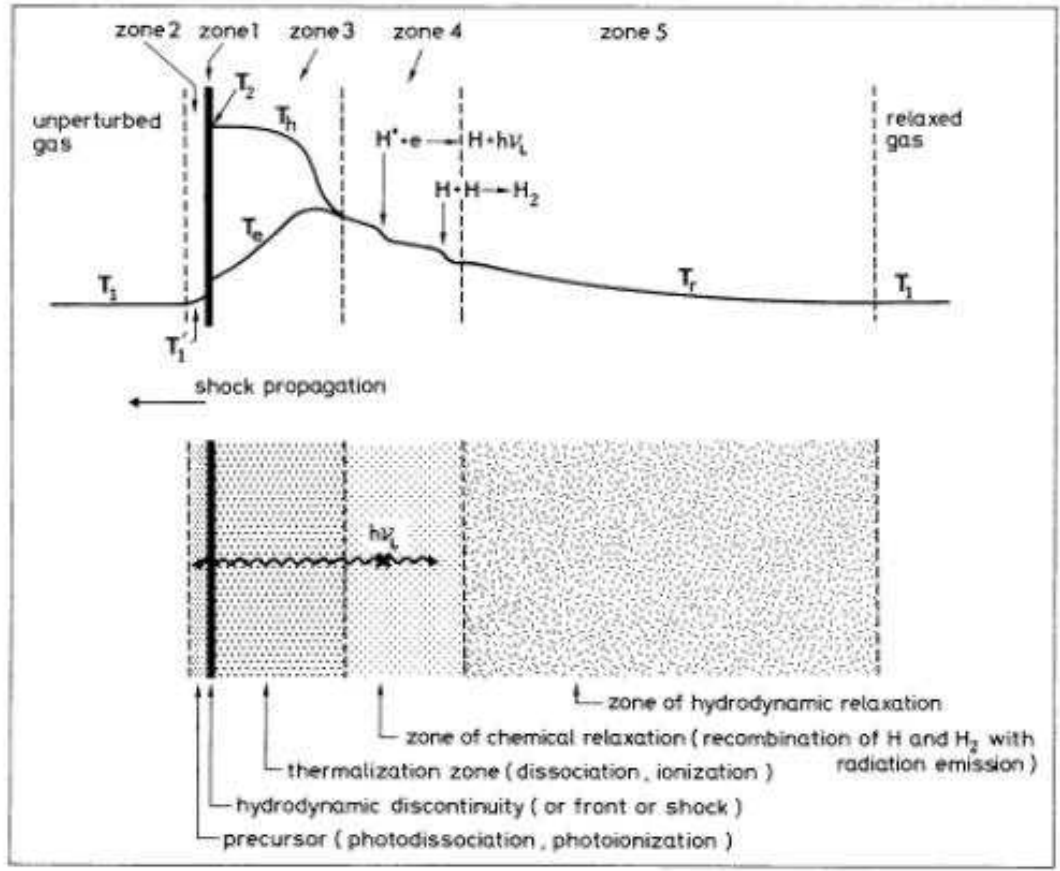}
\vskip0pt
\caption{Shock wave schematic structure  showing several principal zones with different physical characteristics.  }
\label{structure}
\end{figure}

Figure\,\ref{structure} shows an illustration of a shock wave propagating structure, as described 
by Zel'dovich and Raiser (1960) in a cool stellar medium. The shock structure can be subdivided into several principal
zones with different physical characteristics. For large Mach numbers --strong shocks-- the higher collision 
rate caused by the high increase of the temperature causes a thermal dissociation and even ionization of chemical elements 
(in the thermalization zone, zone\,3). In this case, the radiative recombination (in the chemical relaxation zone named the cooling 
wake, zone\,4) produces a photon energy 
propagating much faster than the shock wave (photo-dissociation or photo-ionization zone called precursor,  zone\,2). The shock front (zone\,1), 
the zone where the average parameter discontinuities of the chemical particles steeply change (density, velocity 
and temperature), is much thinner than the precursor and gives direct information on the shock velocity and strength. 
On the other hand, the shock wake is the most important zone, because it is directly 
connected to the formation of the observed emission lines and their interpretation. It is partially transparent both 
in the Balmer and higher continua and in weak spectral lines. The whole shock structure is very sensitive to the degree of 
ionization in the precursor, especially if it was initially weak. At high Mach numbers ($\geq$\,10), the shock is fully 
dominated by radiative phenomena. The coupling between hydrodynamic and radiative fields is high and can completely dominate 
the shock structure through the transfer equation, which is non-local (more details in Chadid et al. 2008).
The dynamics of the atmosphere of RR\,Lyrae stars can be characterized by the passage of two shock waves. The {\it main shock} , at the hump, has higher intensity
and rises into the atmosphere. It is due to the $\kappa$ mechanism. The {\it early shock}, at the bump, is the consequence of the ballistic motion of the atmosphere.    
According to the observations (Table\,1), the intensity of the main shock is very high and can generate
a total ionization of some neutral elements (hypersonic shock, Chadid et al. 2008), in addition to emission, double emission line structure and absorption line doubling 
phenomena (Preston et al. 1965; Chadid \& Gillet 1996). The intensity of the early shock is high enough to generate only an emission in Balmer lines (Gillet \& Crowe 1988).

\begin{flushleft}
\begin{deluxetable*}{rccc}
\tabletypesize{\normalsize}
\tablecaption{Brief history of the detection of the direct signatures of shock waves propagating in the atmosphere of RR\,lyrae stars. }
\tablewidth{0pt}
\tablehead{ \\ \colhead{Shock observational signatures} &\colhead{Pulsation Phases} &  \colhead{First detection by} \\
}

\startdata
Calcium emission lines & hump & Preston \& paczynski (1964)\\
Balmer emission lines & hump& Struve (1947) and Preston \& paczynski (1964)\\
Balmer emission lines & bump & Gillet \& crowe (1988)\\
Balmer emission doubling& hump& Gillet \& crowe (1988) and Chadid et al. (2008)\\
Balmer absorption doubling  &hump & Sanford (1949) and Preston et al .(1965) \\
Metallic line broadening & hump & Chadid \& Gillet (1996b)\\
Metallic absorption doubling  &hump& Chadid \& Gillet (1996a)\\
Neutral metallic lines disappearance & hump& Chadid et al. (2008)\\
Helium absorption lines &hump & Preston (2009) \\
Helium absorption doubling &hump & Preston (2009) \\
Helium emission lines & hump & Preston (2009) \\

\enddata
\end{deluxetable*}
\end{flushleft}

\subsubsection{Hydrogen profiles}
\begin{figure}
\centering
\plotone{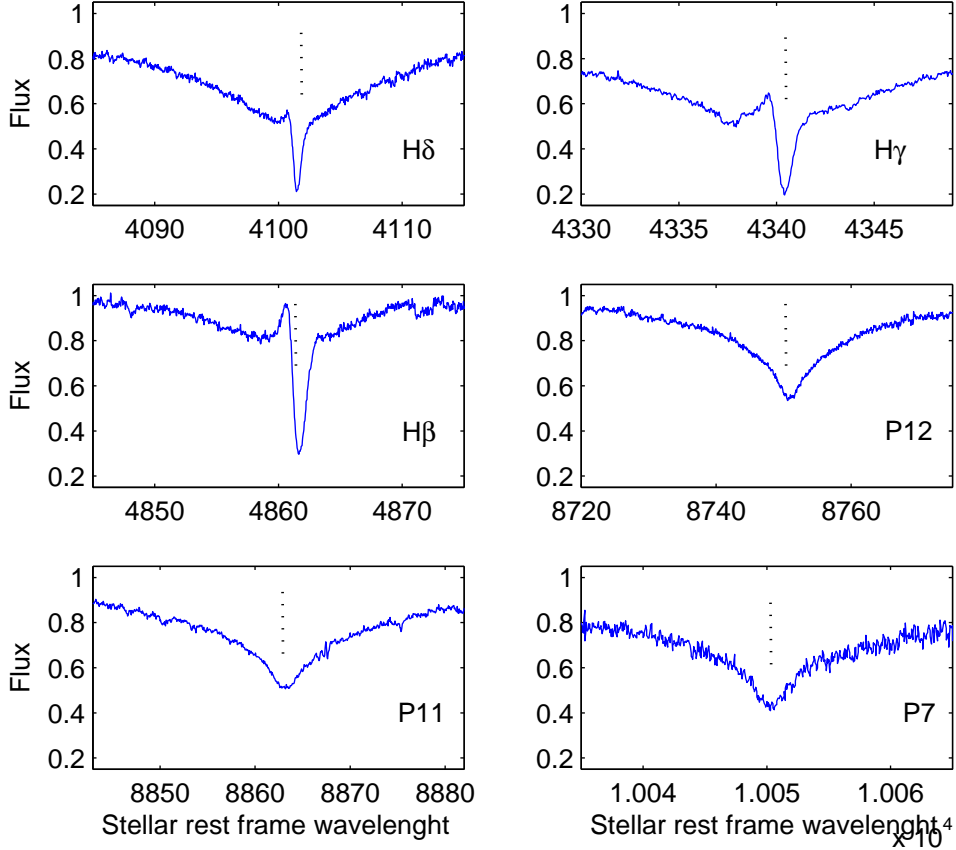}
\vskip0pt
\caption{Hydrogen lines at the maximum emission phase of S\,Arae, using UVES. The Balmer emission decreases from  H\,$\beta$ to H\,$\delta$. No Paschen 
emission line is detected.}
\label{pashen}
\end{figure}

Figure\,\ref{pashen} shows the hydrogen lines at the maximum of the emission phase of S\,Arae observed by UVES. The Balmer lines present a strong blue shifted emission with a smaller but a significant 
red shifted one associated with the hump (Chadid et al. 2008), observed during a time interval of  5\,\% of the pulsation period and centered at phase 0.95. The  H\,$\alpha$ emission is the 
strongest one and the Balmer emission decreases from  H\,$\alpha$ to H\,$\delta$. No Paschen emission line is detected. The emission is formed in the zone of the cooling wake (zone 4 in 
Figure\,\ref{structure}). The H\,$\alpha$ line is one of the strongest lines and formed within the whole atmosphere. The wings are formed deep in 
the atmosphere while the core is formed higher. The variations of the  H\,$\alpha$ line provide direct information about the dynamics of  higher and lower atmosphere. 
H\,$\delta$ is weaker than  H\,$\alpha$ and is formed deeper in the atmosphere. The Balmer double emission structure (consequence of a combination of a 
single large emission and a strong photospheric absorption) has been reported recently in  H\,$\alpha$ and  H\,$\beta$ and it is a consequence of a high shock front 
velocity (Chadid et al. 2008). None have been detected in H\,$\gamma$ and  H\,$\delta$. This means that the  H\,$\delta$  and the Pashen lines are less affected by 
the shock waves propagating in the atmosphere. 

\vfill\eject

\subsection{Evolution of Helium lines}
\begin{figure}
\centering
\plotone{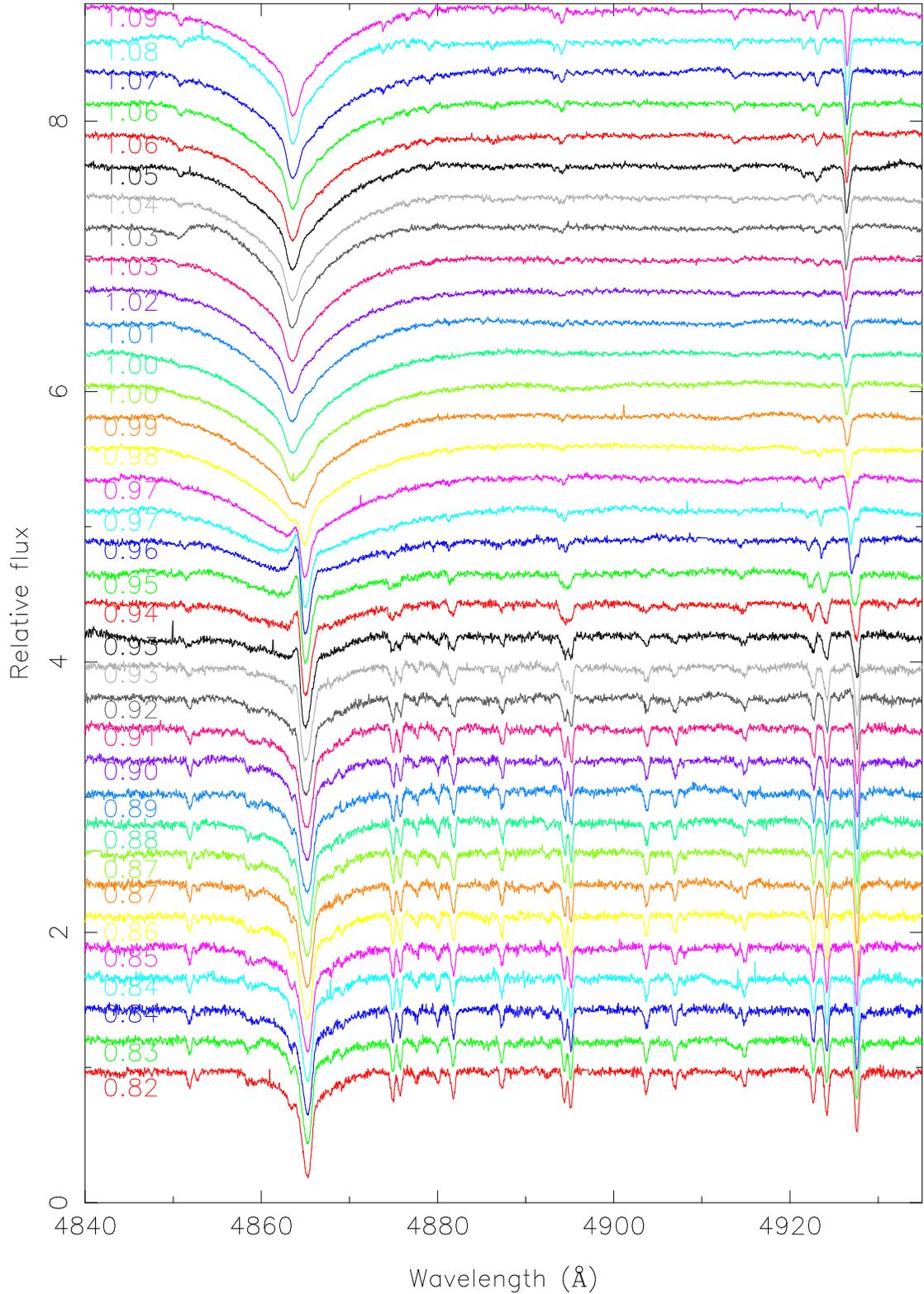}
\vskip0pt
\caption{UVES observed heliocentric lines of S\,Ara, associated with the hump. The phase is labeled in the left corner.}
\label{full}
\end{figure}

Chadid et al. (2008) put into evidence the existence of hypersonic shock waves propagating across the  atmosphere of S\,Arae. They detected a neutral metallic 
line disappearance during the hump (see also Figure\,\ref{full}). They pointed out that the shock energy is high enough to ionize the atoms, forming the layer crossed by the shock wave. 

Preston (2009) detected the neutral helium lines --emission and absorption-- in spectra of eleven RR\,ab stars during rising light, confirming 
the shock Schwarzschild model in RR\,ab stars. He reported the detection of the single ionized helium emission as well in some RR\,ab of his survey.

Upon careful inspection of UVES and HARPS spectra of S\,Arae, I found that the spectra of the hypersonic star S\,Arae reveal, as well, many emission and absorption helium lines. 
This confirms Preston's detection of helium lines (Preston 2009). Table\,2 shows the helium list  detected in S\,Arae. The He\,
5875.62\AA\ is one of the 
strongest Helium lines in the visible region. 

\begin{flushleft}
\begin{deluxetable*}{rccc}
\tabletypesize{\normalsize}
\tablecaption{Detected Helium lines during rising light of the Blazhko and hypersonic star S\,Ara. }
\tablewidth{0pt}
\tablehead{ \\ \colhead{WL ($\AA$)} &\colhead{Element}   & \colhead{Instrument} \\
}

\startdata
4009.27 & He\,I& UVES  \\
4026.45 & He\,I& UVES  \\
4471.48 &He\,I& UVES \\
4685.68 & He\,I\rlap{I}& UVES\\
5875.62 & He\,I& HARPS\\
6678.15 & He\,I& HARPS\\
7065.19 & He\,I& UVES\\
7281.48 & He\,I& UVES\\
\enddata
\end{deluxetable*}
\end{flushleft}

\vskip-24pt

\begin{figure*}
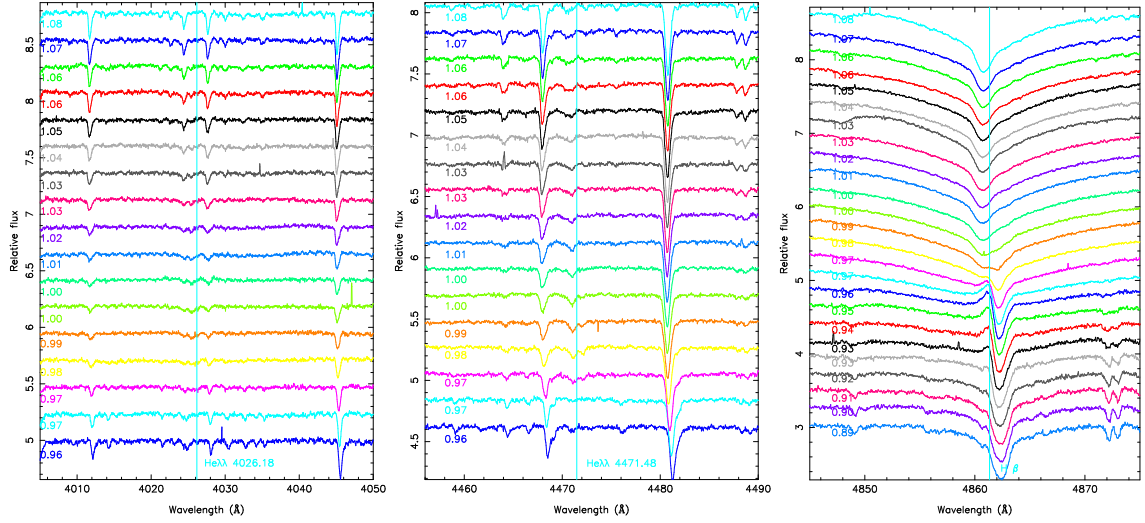

\includegraphics[width=5cm]{he4026.ps}
\includegraphics[width=5cm]{he4471.ps}
\includegraphics[width=4.8cm]{beta.ps}
\vskip0pt
\caption{UVES profile variations of He\,I 4026.45\AA , He\,I 4471.48\AA\ and H\,$\beta$ observed for S\,Arae at phases  labeled in the left corner.
The vertical line give the position of the laboratory line wavelength.}
\label{beta}
\end{figure*}

\begin{figure*}
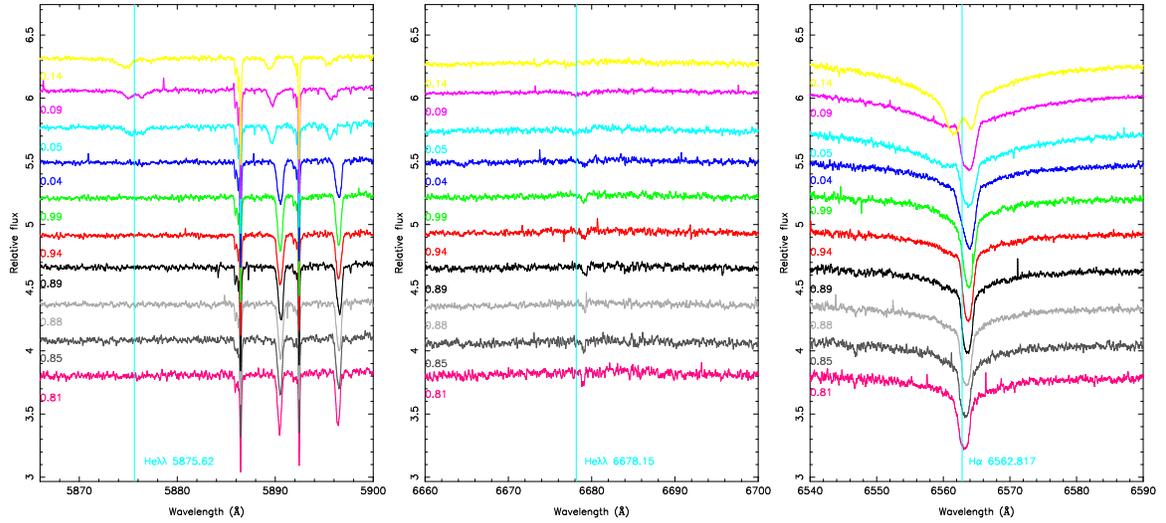

\includegraphics[width=5cm]{he5876.ps}
\includegraphics[width=5cm]{he6678.ps}
\includegraphics[width=5cm]{halpha.ps}
\vskip0pt
\caption{HARPS profile variations of He\,I 5875.62\AA\ , He\,I 6678.15\AA\ and H\,$\alpha$ observed for S\,Arae at phases  labeled in the left corner.
The vertical line give the position of the laboratory line wavelength.}
\label{alpha}
\end{figure*}

From the S\,Arae observations (Section \ref{obs}), only the HARPS echelle orders contain the   
HeI\, 
5875.62 \AA\ line. Figure\,\ref{alpha} shows the evolution
of    HeI\, 5875.62\AA\ and    HeI\, 6678.15\AA\   with H\,$\alpha$ using HARPS with time resolution of 10 to 15\,min.  Figure\,\ref{beta} shows   the evolution
of    HeI\, 5875.62\AA\ and    HeI\, 6678.15\AA\   with H\,$\beta$ using UVES with an exposure time of 5\,min.  In this work, the evolution of the neutral helium profile is consistent 
with Preston (2009) results, and confirm, in a general way,  the shock model proposed by Schwarzschild: first only a pure emission  appears at strongest Hydrogen  emission phase (0.96). 
This emission sometimes is followed by the appearance of a first He component --red shifted-- and 
then by a second He component --blue shifted--  which strengthens with advancing phase and reflecting the line doubling phenomenon like metallic and Balmer line doubling phenomena (Chadid et al. 2008). 
Both helium components (red and blue shifted) gradually weaken and disappear. The  lifetime of the He absorption line mechanism is nearly equal to  that of the neutral metallic line disappearance 
phenomenon (Chadid et al. 2008). Both of them occur around the maximum luminosity  and during a phase interval of 8\,\% of the pulsation period. 

\begin{figure}
\centering
\includegraphics[width=10cm]{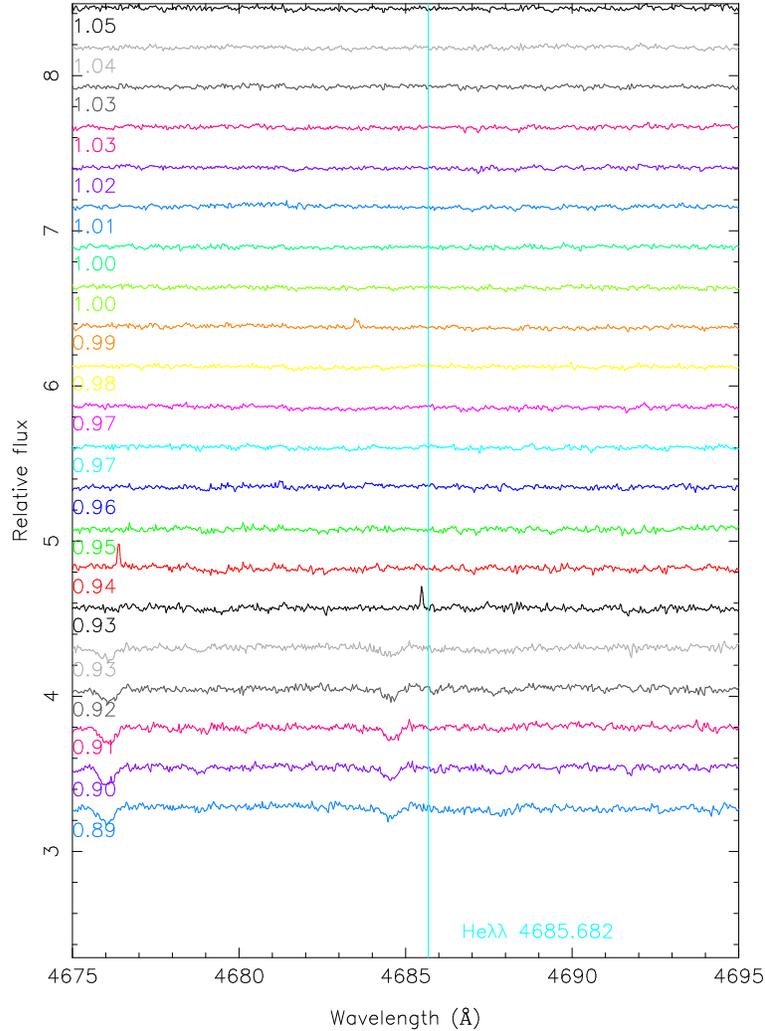}
\vskip0pt
\caption{UVES profile variations of He\,II 4685.68\AA\ observed for S\,Arae at phases labeled in the left corner.
The vertical line give the position of the laboratory line wavelength.}
\label{heii}
\end{figure}

Like hydrogen emission, helium emission lines should be formed just behind the shock front, in a zone of  cooling wake, due to the de--excitation of the neutral helium atom. 
The helium emissions require greater excitation energy than hydrogen lines. Thus, their detection indicates that the shock velocity is strongly supersonic and the conditions within the 
shock wake are extreme. This is consistent with the hypersonic shock detection in S\,Arae. I recall that the minimum energy required for a high--excitation of a neutral helium atom is 
19.74\,eV, equivalent to the excitation threshold temperature of 229,082\,K. The temporal evolution of the helium doubling phenomenon is shown to be in general agreement with the classical 
Schwarzschild mechanism. A new shock wave crosses the deepest atmospheric layers when the ballistic motion induced by the  previous shock is far from being relaxed. The shock at the beginning 
expands with the atmosphere, crossing the Helium line forming region with a velocity higher than the layers and a temperature equal to the temperature of excitation of helium atom. 
The shock reaches the  layers of the helium region at phase 0.96 and produces the double components (blue and red shifted), the red shifted component disappearing more and more. 
Finally, the shock leaves the helium region completely at phase 1.05 when a total disappearance of helium absorption is observed. Figure\,\ref{heii} shows a weak emission of the single ionized helium 
line HeII\, 4685.68\AA , a similar detection has been reported by Preston (2011) in AS\,Vir, UV\,Oct and V1645\,Sgr. This means that, the shock temperature is high enough to ionize even the neutral helium atom, and 
reaches  a temperature value of 285,367\,K (the temperature required  for an ionization of a neutral helium atom and equivalent to a minimum energy of 24.59\,eV). This means that the shock wave in RR\,lyrae stars 
S\,Arae, AS\,Vir, UV\,Oct and V1645\,Sgr is extremely strong --a hypersonic shock-- and can reach a Mach number greater than 30.

\section{Magnetic field in the Blazhko star RV\,UMa}
\subsection{High resolution spectro-polarimetric observations}
Using the fiber–fed echelle spectro-polarimeter NARVAL (Auriere 2003) at the 2-m Telescope Bernard Lyot (TBL) of Pic du 
Midi Observatory, I carried out, on February 27 and 28th, 2009, high precision (median $\sigma_{\rm B}\sim 35$~G) 
longitudinal magnetic field measurements of the Blazhko RR\,Lyrae star RV\,UMa, which is an RR\,ab star with a large amplitude, 
(V--magnitude is between 10.11 and 11.23), a
 pulsation period of 0.468 days and a Blazhko cycle of 90 days (Preston \& Spinrad 1967). The Stokes V spectra were obtained 
at various pulsation phases and have been reduced using the libre-ESPRIT data reduction package (Donati et al 1997). Total exposure 
time for a full Stokes $V$ exposure (4 sub-exposures) was typically 60 min, leading to signal-to-noise ratio (S/N) in the range of 
700--1000. The spectra were normalized using an automated code with a polynomial fit of 5 or 6 degrees. To
obtain a longitudinal field measurement, a Least-Squares Deconvolution (LSD; Donati et al.1997)  multiline analysis procedure was 
employed to extract high-precision mean Stokes $I$ and $V$ profiles from each spectrum.  The line mask employed for
extraction of the RV\,UMa LSD profiles was computed using a VALD {\tt extract stellar} line list and a solar abundance table.  
The RV\,UMa line mask contained 750 metallic lines within the spectral range 370--1000\, nm.  The typical S/N of the LSD Stokes 
$V$ profiles is about 3600:1. No significant circular polarization is detected in any  RV\,UMa LSD Stokes $V$ profiles.
   For comparison, I obtained one observation of the prototypical magnetic Ap standard star 
$\alpha^2$\,CVn . The S/N ratio in the $\alpha^2$~CVn Stokes $V$ LSD profiles is typically around 5000:1. In contrast to RV\,UMa, 
significant circular polarization is detected in LSD Stokes $V$ profile of $\alpha^2$~CVn. 

\vfill\eject

\subsection{Longitudinal magnetic field measurements}
The  determination of the longitudinal magnetic field values and uncertainties are based on the use of the formula 
$$ B_{\ell}\,(G) = -2.14 x 10^{11} \frac{\int_{}^{}vV(v)dv}{\lambda gc
\int_{}^{}[Ic - I(v)]dv} $$

\noindent (Chadid et al. 2004) where $g$ is the mean value of the Land\'e factors of all
lines used to construct the LSD profile, $\lambda$ is the mean wavelength and $c$ is the velocity of light. 
The pulsation phase of RV\,UMa has been calculated  using the following ephemerids:\\
HJD(max.light)\,=\,2,\,445075.51100\,+\,0.468060\,E (light),\\ The maximum pulsation corresponding to $\phi$=0.00 .   The inferred values for the  longitudinal magnetic fields of RR Lyr
and $\alpha^2$~CVn are reported in the  journal of observations (Table\,3).

\begin{flushleft}
\begin{deluxetable*}{rccccccc}
\tabletypesize{\normalsize}
\tablecaption{Journal of observations of NARVAL spectro-polarimeter of RV\,UMa at TBL and the longitudinal magnetic field measurements. 
The heliocentric Julian Date is given at mid--exposure. The last column lists the derived longitudinal magnetic field values and their corresponding uncertainties.}
\tablewidth{0pt}
\tablehead{ \\ \colhead{HJD} &\colhead{Object}   & \colhead{Type} &
       \colhead{V$_{\rm (mag)}$} & \colhead{Exp$_{\rm (sec)}$} & \colhead{Rotational or Pulsation Phase} &
  \colhead{$B_\ell \pm \sigma_{\rm B}$}  \\
}

\startdata
2454889.690277 & $\alpha^2$ $\,CVn$  & Ap star   & 3.19 & 120  & 0.902   & \llap{$-$7}00$\pm$26\\
2454889.572910 & RV UMa  & RRab   & 11.30 & 3600  & 0.529    & \llap{$+$}24$\pm$56\\
2454889.614583&RV UMa  & RRab   & 11.30 & 3600  &0.368   & \llap{$-$}15$\pm$20\\
2454889.661111&RV UMa  & RRab   & 11.30 & 3600  &0.718    &\llap{$+$}52$\pm$42\\
2454890.549305&RV UMa  & RRab   & 11.30 & 3600  &  0.615   & \llap{$-$}58$\pm$40\\
2454890.549305&RV UMa  & RRab   & 11.30 & 3600  &0.037    & \llap{$-$}31$\pm$31\\
2454890.636805&RV UMa  & RRab   & 11.30 & 3600  &0.802    & \llap{$-$}50$\pm$22\\

\enddata
\end{deluxetable*}
\end{flushleft}

The longitudinal field measurements of RV\,UMa show values between 15--58 \,G, with 1\,$\sigma$
uncertainties between 20--56\,G,  leading to  a mean longitudinal magnetic field of $B_{\ell} = -13\pm 35$~G. None of 
measurements of  RV\,UMa show a significant detection (the mean $z=|B_\ell/\sigma_{\rm B}|$ parameter is around 1.35).

\subsection{Constraints on the magnetic field}
\begin{figure}
\centering
\plotone{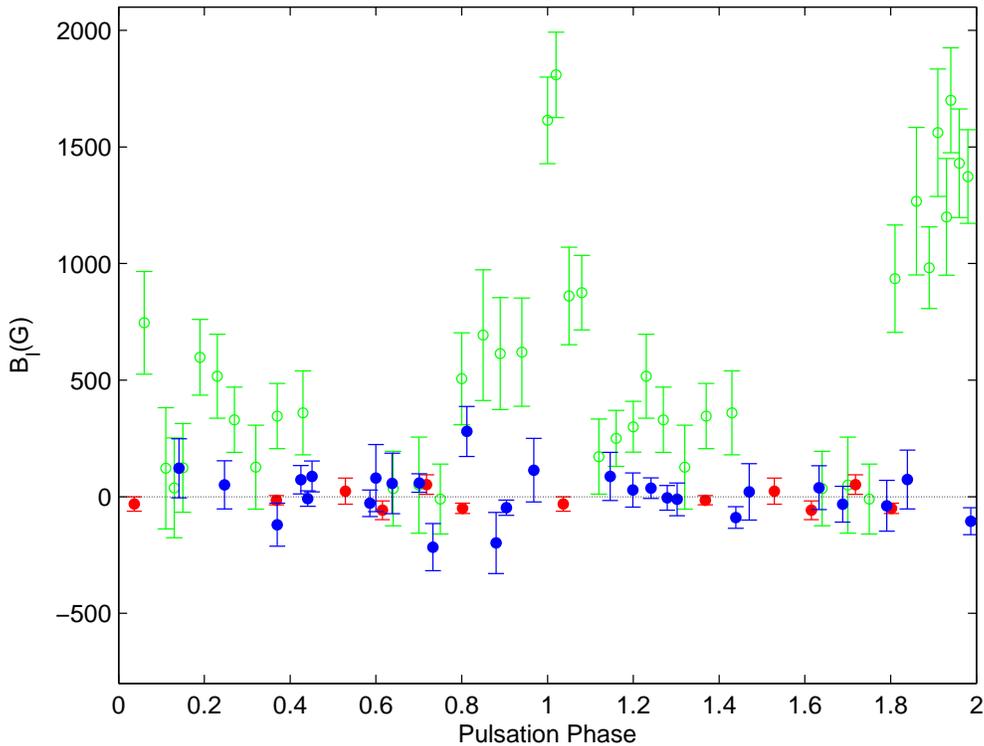}
\vskip0pt
\caption{Comparison of NARVAL longitudinal magnetic field measurements (RED solid circles) of RV\,UMa versus pulsation phase
   with the MuSiCoS longitudinal magnetic field of RR\,Lyr (Chadid et al. 2004)(BLUE solid circles)  and  magnetic field
   variation of RR\,Lyr reported by Romanov et al. 1987 in 25-28 September 1982 (GREEN open circles), phased according to the pulsational
 stellar ephemeris.}
\label{mag}
\end{figure}

\begin{figure}
\centering
\plotone{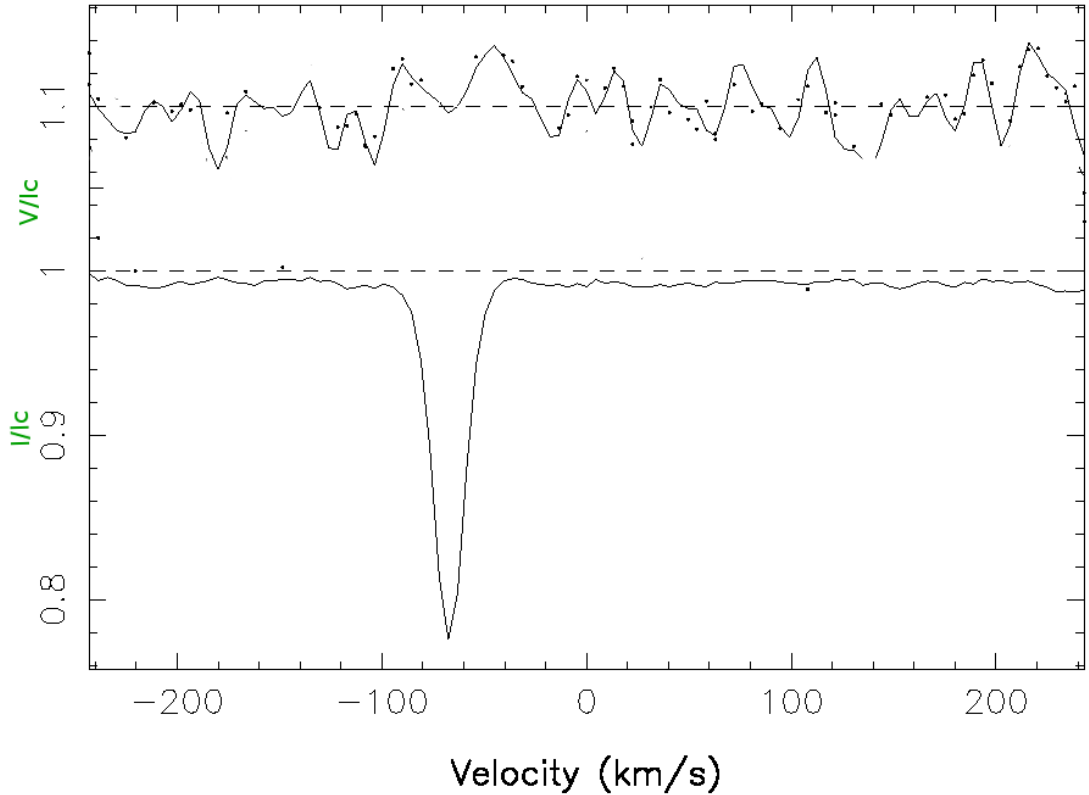}
\vskip0pt
\caption{LSD unpolarized Stokes I (lower curve) and circularly polarized Stokes V (upper curve) profiles of RV\,UMa. 
obtained on 28 February 2009 corresponding to minimum pulsation phase ($\phi$ \, $\sim$ \,0.50). the heliocentric radial velocity 
value is -69.23\,km/s. No significant Zeeman signature is detected.}
\label{stokes}
\end{figure}

 According to Romanov et al. (1987, 1994) the Blazhko star RR\, Lyr exhibits a longitudinal magnetic field which varies 
approximately sinusoidally, with a maximum value of $\sim 1800$~G and an amplitude of the field variation
 of $\sim 1000$~G , when phased according to the pulsation period of $0.567$ days. By using high precision circular polarization over 4--year cycle, 
Chadid et al. (2004) reported that RR\,Lyr is a bona fide non--magnetic star, a result inconsistent with the magnetic field detected
by Romanov et al. (1987, 1994).  Chadid et al. explained that the strong conflict with their measurements and those of Romanov et al 
is due to  the limitations of the use of photographic plates resulting in underestimated error bars and  distortions of of the 
metallic lines caused by the shock waves propagating during certain pulsation phases (phase $\phi$\,=\,0.95 where Romanov et al. measure a magnetic 
field  larger than the typical sigma of Chadid et al.).  Figure \ref{mag} shows a comparison of 
RV\,UMa longitudinal field measurements with those of chadid et al. (2004) and  Romanov et al. (1987). It is clear that  RV\,UMa 
results are consistent with our previous results (Chadid et al. 2004) and point out that if a dipole field is present, its strength must be extremely 
smaller than the 1.8\,kG field of Romanov et al. None of RV\,UMa LSD Stokse $V$ profiles (Figure\,\ref{stokes}) put into evidence of any circular polarization in the mean Stokes
profiles of  RV\,UMa over one pulsation cycle. 

The RV\,UMa observations follow the line of results obtained by Preston (1967) and Chadid et al (2004) for RR\,Lyr itself and reveal a serious
 challenge to the magnetic models for explaining the Blazhko effect.

\vfill\eject

\section{Stothers' explanation of the Blazhko effect}
Our results on the magnetic field of RR\,Lyrae stars with different pulsational properties imply that the oblique--dipole rotator model (Balazs-Detre 1964) is not able to determine the 
relevant mechanism at the origin of the Blazhko effect. Stothers (2006, 2010)  suggested that the turbulent/rotational dynamo mechanism generates, in some RR\,Lyrae stars, the magnetic 
fields that grow over the Blazhko cycle and suppress turbulent convection, and in turn small changes in 
the period of the fundamental radial mode.  The physical mechanism suggested by Stothers appears very 
promising, but in order to fully check this hypothesis more detailed predictions about the key observables such as amplitude, phase and period modulations are required. Recently, 
Chadid et al. (2011) detected in the 0.567 days variable CoRoT ID 0105288363, significant cycle--to--cycle changes in the Blazhko modulation, 
which appear to be analogous to those predicted by Stothers.  They  discussed the  clear correlations between the phase and the amplitude of the 
bump, and the  skewness and acuteness of the light curve during different Blazhko cycles. They found that these quantities are strongly anti-correlated with the fundamental pulsation period. 
This provides a strong support to the slow convective cycle model suggested by Stothers. 

However, how does slow convective cycle happen in Blazhko hypersonic atmospheres?

Struve (1947) discovered the occurrence of transitory Hydrogen emission in RR\,Lyr itself and Struve and Blaauw (1948) showed that its intensity varies in the 41-day cycle (Blazhko cycle). 
Preston, Smak and Paczynski (1965) found that the visibility of doubling of the Hydrogen absorption components is correlated with the strength of the Hydrogen emission in the 41--day cycle 
and also that the doubling of  the higher members of the Balmer series always precedes that of the lower ones. From the data of these two last studies, it appears that the observed average radial 
velocity over one pulsation period is also a function of the 41--day cycle. Preston, Smak and Paczynski (1965) interpreted these phenomena 
as a consequence of a variation of the critical level of the shock-wave formation region during the 41--day cycle. The doubling phenomenon of the metallic absorption lines is the predominant 
contribution of the considerable and very narrow FWHM increase observed in RR Lyr (Chadid and Gillet 1996b). This latter could provide a point of contact between observation and Stothers' hypothesis
(as mentioned already by Preston 2011). Moreover, the spectroscopic work of Chadid \& Gillet (1997) reveals that the amplitude of shock waves is greatly correlated with the Blazhko modulation. 
Thus, the stothers's proposal, slow convective cycles in the envelopes of RR\,lyrae stars (Stothers 2001), seems to be not enough to explain the Blazhko modulation of RR\,Lyrae 
hypersonic atmospheres.

\section{Conclusion}
``{\it It takes a village to raise a child, as H. Clinton (2006) said in her book... There is an analogy for this view in the world of Science}'' said George Preston. 

I started my spectroscopic works in Observatoire de Haute Provence (OHP). I learned from the oldest and most aged spectroscopic school of OHP and I had a great chance to meet the  
most experienced photographers such as Fehrenbach, Wallerstein, Woltjer, Schatzman, Barbier etc. They took me towards the oldest photographic spectra and   works of George Preston on the
atmosphere  of RR\,lyrae stars. 
Now, I am sharing my experiences and skills with George Preston and together (face to face) we are exploring  the most shocked atmospheres: atmospheres of RR\,Lyrae stars. 
Using the newest and most modern technologies like CCDs.

  Happy  birthday George and thank you for inviting me  to collaborate with you. Thank you for the stay in Carnegie Observatories and for introducing me to your ``{\it village}''. It is really 
a ``{nice village}''.

\end{document}